\documentclass[journal = jpclcd, manuscript=letter]{achemso}

\usepackage{chemformula} 
\usepackage[T1]{fontenc} 

\author{Juan B. P\'erez-S\'anchez}
\email{jperezsa@ucsd.edu}
\author{Joel Yuen-Zhou}
\email{joelyuen@ucsd.edu}
\affiliation[UCSD]
{Department of Chemistry and Biochemistry, University of California San Diego, La
Jolla, California 92093}
\title[paper1]{Polariton assisted down-conversion of photons via nonadiabatic molecular dynamics: a molecular dynamical Casimir effect}


\begin{document}

\begin{tocentry}

\includegraphics[width=5.1cm,height=5.1cm]{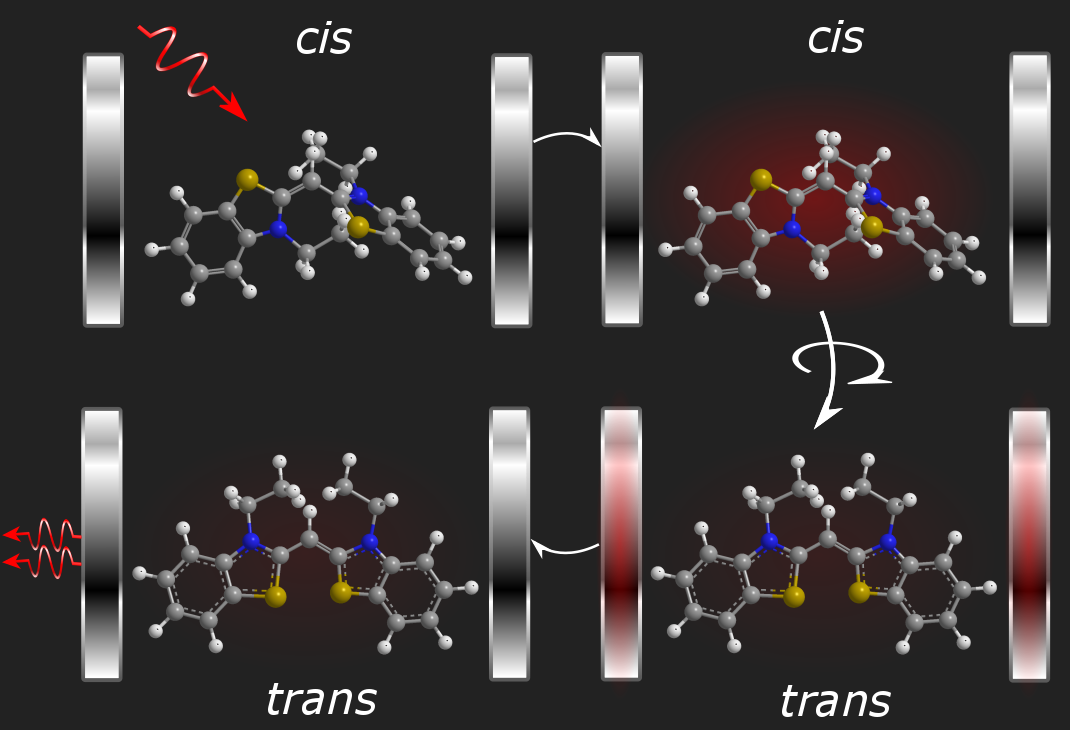}

\end{tocentry}

\begin{abstract}
Quantum dynamics of the photoisomerization of a single 3,3'-diethyl-2,2'-thiacynine iodide molecule embedded in an optical microcavity was theoretically studied. The molecule was coupled to a single cavity mode via the quantum Rabi Hamiltonian, and the corresponding time-dependent Schr\"odinger equation starting with a purely molecular excitation was solved using the Multiconfigurational Time-Dependent Hartree Method (MCTDH). We show that, for single-molecule strong coupling with the photon mode, nonadiabatic molecular dynamics produces mixing of polariton manifolds with differing number of excitations, without the need of counterrotating light-matter coupling terms. As a consequence, an electronic excitation of the molecule at {\it cis} configuration leads to the generation of photon pairs in the {\it trans} configuration upon isomerization. Conditions for this phenomenon to be operating in the collective strong light-matter coupling regime are discussed and found to be unfeasible for the present system, based on simulations of two molecules inside the microcavity. Yet, our finding suggests a new mechanism that, without ultrastrong coupling, achieves photon down-conversion by exploiting the emergent molecular dynamics arising in polaritonic architectures.
\end{abstract}


Most photochemical and spectroscopic processes studied at present correspond to the so-called weak light-matter coupling regime. In the latter, the effect of light is merely to produce transitions between molecular eigenstates. However, using confined electromagnetic media, such as those in optical microcavities \cite{vahala} or nanostructures \cite{Vasa}, it has been possible to reach new regimes in which the light-matter interaction energy is comparable to electronic or vibrational energies, so that light has to be explicitly considered beyond a perturbative treatment \cite{Nori,Raphael}. The new hybrid light-matter states in strong coupling (SC) and ultra-strong coupling (USC) regimes are called polaritons. These regimes do not necessarily require high intensity lasers, but occur despite the cavity modes being in their vacua or low-lying excitations.

In the case of a single molecule interacting with a single cavity mode, SC and USC regimes are commonly described by the Rabi model \cite{Chikkaraddy,Nori,Santhosh} where the matter part is taken to be a two-level system with no internal structure, an insufficient description to study molecular processes where nuclear dynamics plays a major role \cite{Prineha}. Theories that take into account the correlated nuclear-electronic-photonic dynamics have been recently developed to account for the rovibrational structure of molecules \cite{Rubio2,Feist2,Maitra}. In these new frameworks, molecular-photonic dynamics are described in dressed or polaritonic Potential Energy Surfaces (PESs), and are governed by novel features such as light-induced avoided crossings (LIACs) and light-induced conical intersections (LICIs) \cite{Moiseyev,Kowalewski,Kowalewsk,Gabor2,Cederbaum}. 

The idea of using light to drive chemical reactions with high selectivity has been one of the dreams of chemistry since the invention of the laser, and the ability to tune the characteristics (polarization, mode volume, spatial profile) of the optical modes of a confined photonic media seems to offer new and versatile control knobs to manipulate molecular properties on demand, without invoking costly and time-consuming synthetic modifications \cite{Ebbesen,KenaCohen}. Indeed, most recent works focus on using strong light-matter coupling to change molecular processes such as photodissociation \cite{Triana,Vendrell,Kowalewski3}, photoisomerization \cite{Huo,Feist,Feist1,Feist2,Groenhof,Persico}, and charge and energy transfer \cite{Groenhof,Matt}. In this paper we focus on a less addressed complementary question: can the emergent molecular dynamics under SC be harnessed for photonic applications? By theoretically studying the photoisomerization of a single 3,3'-diethyl-2,2'-thiacynine iodide molecule that strongly interacts with a cavity, we find that for specific cavity frequencies and sufficiently strong couplings, molecular photoexcitation into an electronic excited state can produce the emission of two photons of a lower frequency via the cavity after isomerization, thus offering a new mechanism for photonic down-conversion using molecular polaritons. This phenomenon provides a molecular version of the dynamical Casimir (DC) effect, where photon pair creation arises from nonadiabatic  modulation of the electromagnetic vacuum. As we shall show, molecular nonadiabatic effects mix states with diferent excitation numbers, without the need of the usually invoked counterrotating light-matter coupling terms which are relevant in the standard realizations of the DC effect, which operate under USC conditions \cite{Nori,Stassi,Stassii}.

\begin{center}
\begin{figure}
\includegraphics[scale=0.38]{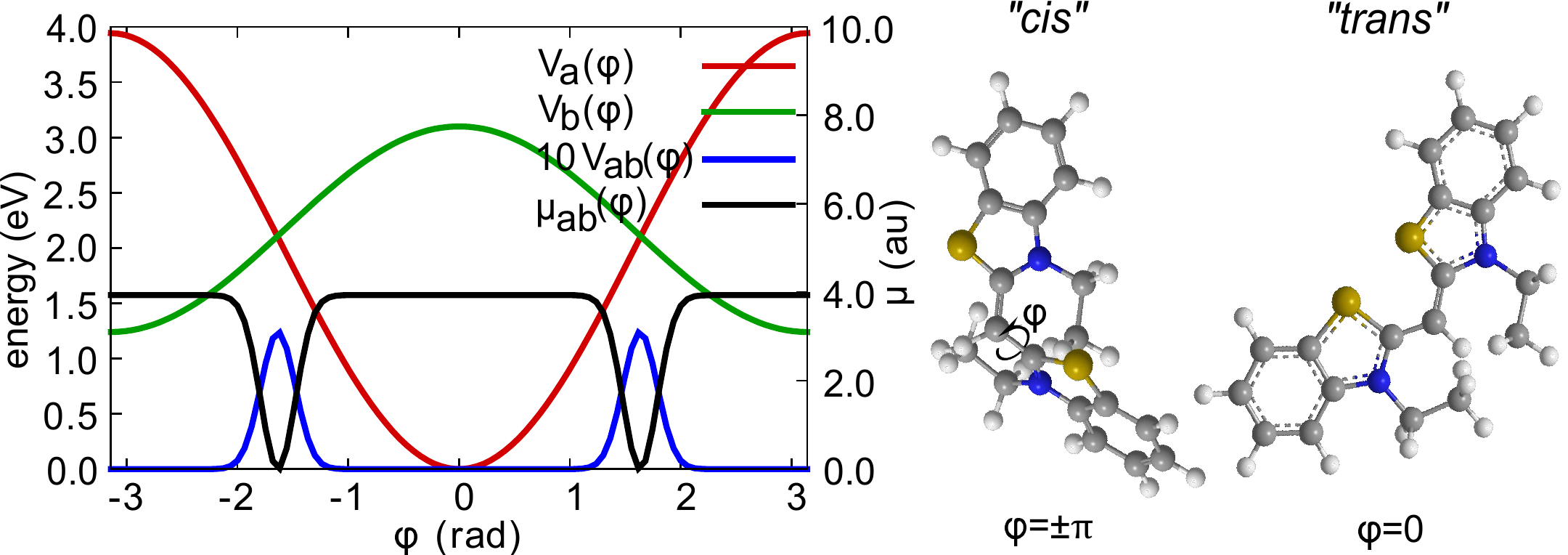}
\caption{Potential energy surfaces $V_{a}(\phi)$ and $V_{b}(\phi)$, diabatic coupling $V_{ab}(\phi)$, and transition dipole moment $\mu_{ab}(\phi)$ for 3,3'-diethyl-2,2'-thiacynine iodide; parameters from Ref.~\citenum{Brumer}. {\it Cis} and {\it trans} configurations are located around $\phi=\pm\pi$ and $\phi=0$ respectively. Diabatic coupling $V_{ab}(\phi)$ generates molecular avoided crossings near $\phi=\pm 1.63\ $rad.}
\label{potentials}
\end{figure}
\end{center}

To begin with, the bare molecular Hamiltonian is given by ($\hbar=1$) $\hat{H}_{mol}= \hat{T}_{N} + \hat{H}_{el}(\phi)$, where $\hat{T}_{N}=-\frac{1}{2 m}\frac{\partial^{2}}{\partial \phi^{2}}$,

\begin{equation}
\hat{H}_{el}(\phi)=\begin{pmatrix}
 V_{a}(\phi) & V_{ab}(\phi)\\ 
 V_{ab}(\phi) & V_{b}(\phi)
\end{pmatrix},
\end{equation} $\phi$ is the torsional angle of the molecule (reaction coordinate), $V_{a}(\phi)$ and $V_{b}(\phi)$ are {\it diabatic} PESs, and $V_{ab}(\phi)$ is the diabatic coupling, responsible to produce transitions between the electronic states $|a\rangle$ and $|b\rangle$. Diagonalization of $\hat{H}_{el}$ as a function of $\phi$ produces adiabatic states of low energy $|g\rangle$ and high energy $|e\rangle$. These purely molecular quantities can be determined by quantum chemistry calculations and spectroscopic measurements. In this work, we take these properties from a previous model parametrized by Hoki and Brumer (see Figure \ref{potentials})\cite{Brumer}.

On the other hand, the Hamiltonian of the cavity mode is given by 

\begin{equation}
\hat{H}_{cav}=\omega_{c}(\hat{a}^{\dagger}\hat{a}+1/2),
\end{equation} where $\omega_{c}$ is the cavity frequency, and $\hat{a}$ is the photon annihilation operator. The Hamiltonian of the cavity-molecule system is the sum of those corresponding to the molecule, the cavity mode, and the interaction between them:

\begin{equation}
\hat{H}= \hat{H}_{mol} + \hat{H}_{cav} + \hat{H}_{I}. \label{totalH}
\end{equation}

In this work, the cavity-molecule coupling is modeled as described in Ref.~\citenum{Kowalewski}, where the photon is coupled to the electronic transition through the molecular transition dipole moment $\mu_{ab}(\phi)$ (see Figure \ref{potentials}). Eq. \ref{totalH} can then be re-expressed as $\hat{H}=\hat{T}_{N}+\hat{H}_{e-p}(\phi)$, with the adiabatic polaritonic BO Hamiltonian given by

\begin{equation}
\hat{H}_{e-p}(\phi)= \hat{H}_{el} + \omega_{c}(\hat{a}^{\dagger}\hat{a}+1/2) +  g(\phi)(\hat{a}^{\dagger}+\hat{a})\hat{\sigma}_{x} 
\end{equation} Here, $g(\phi)=\epsilon \omega_{c} \mu_{ab}(\phi)$ and $\epsilon=1/\sqrt{2V\omega_{c}\epsilon_{0}}$.

\begin{center}
\begin{figure}[ht]
\includegraphics[scale=0.4]{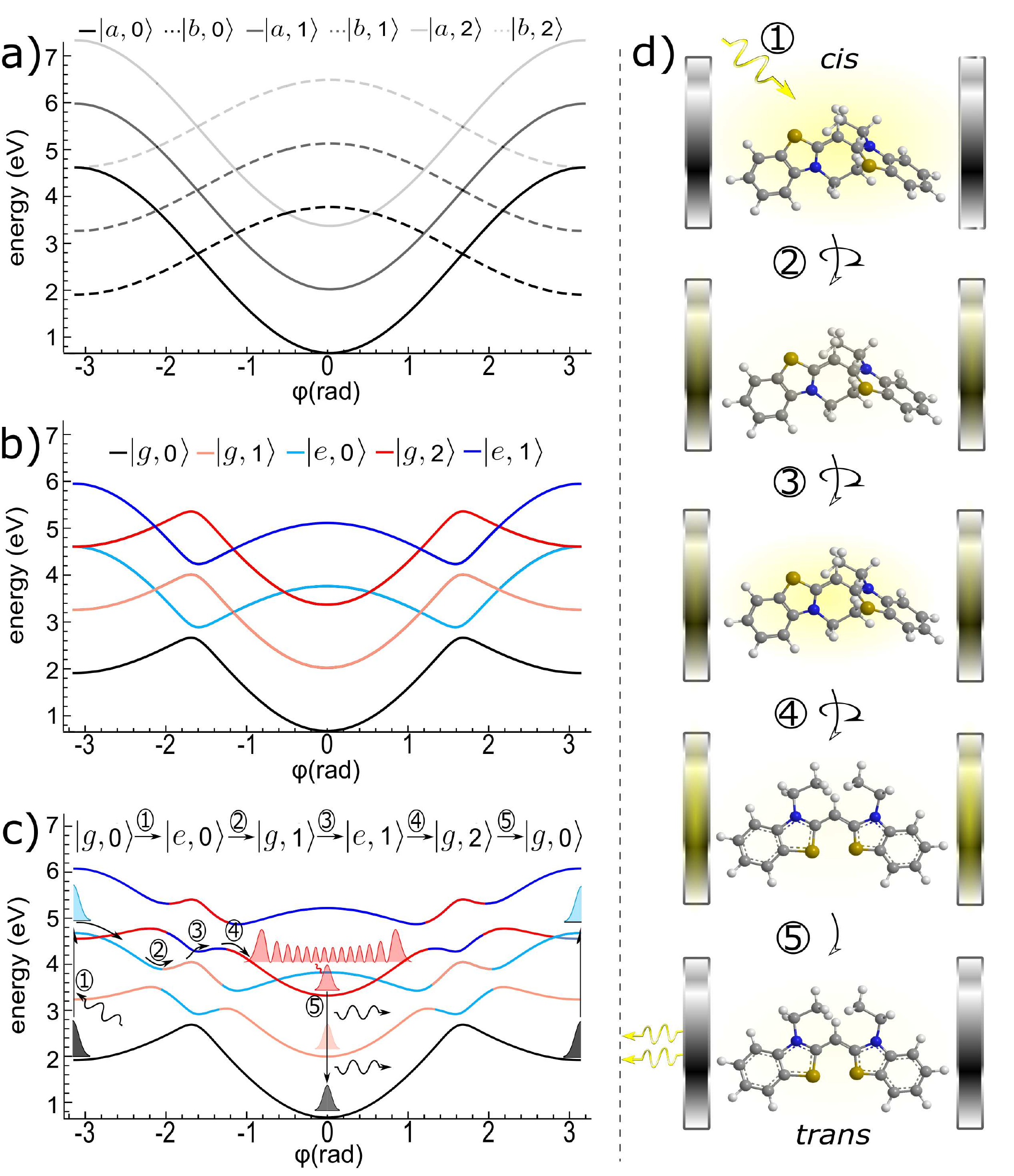}
\caption{Polaritonic PESs for $\omega_{c}=0.5\ \omega_{ab}$. a) Ignoring both light-matter and diabatic couplings. b) Turning on diabatic coupling, generating adiabatic states $|g\rangle$ and $|e\rangle$. c) Turning on both light-matter and diabatic couplings ($\epsilon=0.04$ au $\rightarrow g\approx 0.21$ eV). d) Mechanism of photon down-conversion is shown as a sequence of five steps. Light red: one photon $|g,1\rangle$. Dark red: two photons $|g,2\rangle$. Light blue: one exciton $|e,0\rangle$. Dark blue: one exciton with one photon $|e,1\rangle$. d) Diagrammatic representation of mechanism in c). }
\label{polaritonPES}
\end{figure}
\end{center}

By conveniently using the Fock state basis for the cavity mode, $\hat{H}_{e-p}$ can be diagonalized to obtain adiabatic polaritonic PESs. These are shown in Figure \ref{polaritonPES} for specific values of cavity frequency and light-matter coupling. To establish a reference, the electronic energy gap at the {\it cis} configuration of the molecule is labeled as $\omega_{ab}\equiv V_{a}(\phi=-\pi)-V_{b}(\phi=-\pi)=2.70$ eV. We varied $\omega_{c}$ from 25$\%\ $ to 100$\%\ $ of $\omega_{ab}$, as well as the light-matter scaling parameter $\epsilon$ from 0.01 au to 0.05 au. The latter corresponds to a maximum light-matter coupling $g(\phi)$ ranging from $4\%\ $ to $20\%\ $ of $\omega_{c}$.

Notice that polaritonic PESs (Figure \ref{polaritonPES}c) at the {\it cis} and {\it trans} configurations resemble their counterparts in the absence of light-matter coupling (Figure \ref{polaritonPES}b); this effect is a consequence of the cavity and the molecule being highly off-resonance at those configurations. However, as the torsional angle $\phi$ changes, so does the electronic energy gap, leading to resonances at $\phi=\pm 1.16,\pm 2.14$ rad. Light-matter coupling about these resonances generates LIACs. In this adiabatic polaritonic basis, the kinetic energy operator is not longer diagonal, and it generates nonadiabatic couplings among different polaritonic PES. By analogy with atomic systems or synthetic qubit systems, where different polariton manifolds can be couple through counterrotating terms \cite{Nori,Stassi,Stassii,Ashhab}, coexistence of nonadiabatic and light-matter couplings is enough to mix polaritons with different number of excitations, with the difference that parity in excitation number need not be conserved.

Figures \ref{polaritonPES}c and \ref{polaritonPES}d summarize our proposal to achieve photon down-conversion using molecular polaritons:

\begin{enumerate}
\item Resonant optical excitation of the molecule from state $|g,0\rangle$ to state $|e,0\rangle$. This transition occurs via direct interaction between the molecular dipole and a high frequency photon that is transparent (non-resonant) with respect to the cavity.
\item Adiabatic dynamics across a LIAC converts electronic excitation into a cavity photon ($|e,0\rangle \rightarrow |g,1\rangle$).
\item Nonadiabatic wavepacket dynamics across a molecular avoided crossing converts vibrational energy into electronic energy ($|g,1\rangle \rightarrow |e,1\rangle$). 
\item Adiabatic dynamics across a second LIAC converts electronic excitation into a second cavity photon ($|e,1\rangle \rightarrow |g,2\rangle$).
\item Photons are spontaneously emitted into the electromagnetic bath through the cavity ($|g,2\rangle \rightarrow  |g,0\rangle$).
\end{enumerate}

We emphasize that steps 2 and 4 are possible only if light-matter coupling is strong enough to create a sizable LIAC that favors adiabatic nuclear dynamics. In other words, energy exchange between cavity photon and molecule must be fast compared to the instantaneous nuclear motion at the vicinity of the LIAC. To illustrate our proposal, we numerically solve the Time-Dependent Schr\"odinger Equation using the Multi-Configurational Time-Dependent Hartree method (MCTDH) \cite{mctdh2,mctdh}. Even though we gained much conceptual insight by appealing to a Fock basis for the the cavity mode, we will numerically deal with the latter in quadrature coordinates \cite{Flick},

\begin{equation}
\hat{a}=\sqrt{\frac{\omega_{c}}{2}}\left( \hat{x} + \frac{i}{\omega}\hat{p}\right), 
\end{equation} where $\hat{p}=-i\frac{\partial }{\partial x}$. With these identifications, Eq. (2) can be rewritten as

\begin{align}
&\hat{H}= \hat{T}_{N}+\frac{\hat{p}^{2}}{2}+ \begin{pmatrix}
 V_{a}(\phi) + \frac{1}{2}\omega_{c}^{2}x^{2} & V_{ab}(\phi) + g(\phi) x\\ 
 V_{ab}(\phi) + g(\phi)x & V_{b}(\phi)+ \frac{1}{2}\omega_{c}^{2}x^{2}
\end{pmatrix},
\end{align} where the cavity mode appears as an additional ``vibrational'' coordinate, whose implementation in the MCTDH is straightforward \cite{Kowalewsk}. The wave function is expanded as a linear combination of diabatic electronic states $|k\rangle$:

\begin{equation}
\langle x, \phi | \Psi(t)\rangle = \sum_{k}\psi_{k}(x,\phi,t)|k\rangle, \ \ \textrm{for}\ k=a,b
\end{equation} and the initial state is chosen to represent an impulsive Franck-Condon excitation of the molecule (directly via a high-energy photon that is transparent to the cavity), namely, a product state of the molecular ground state on top of the excited electronic state $|a\rangle$ at the {\it cis} configuration $\varphi_{b}(\phi)$, accompanied by the vacuum state of the cavity mode $\chi(x)$:

\begin{equation}
\langle x, \phi | \Psi(0)\rangle = \varphi_{b}(\phi)\chi(x)|a\rangle.
\end{equation}

To analyze the computational results, we calculate adiabatic populations of electronic and photonic states as a function of time: 

\begin{equation}
P_{\kappa,n}(t)=\langle \Psi(t)| \kappa,n \rangle \langle \kappa,n | \Psi(t)\rangle, \ \ \ \textrm{for} \ \kappa=g,e, \textrm{and } n=0,1,2,....
\end{equation} In this expression, the torsional degree of freedom $\phi$ is traced out.
\begin{center}
\begin{figure}
\includegraphics[scale=0.7]{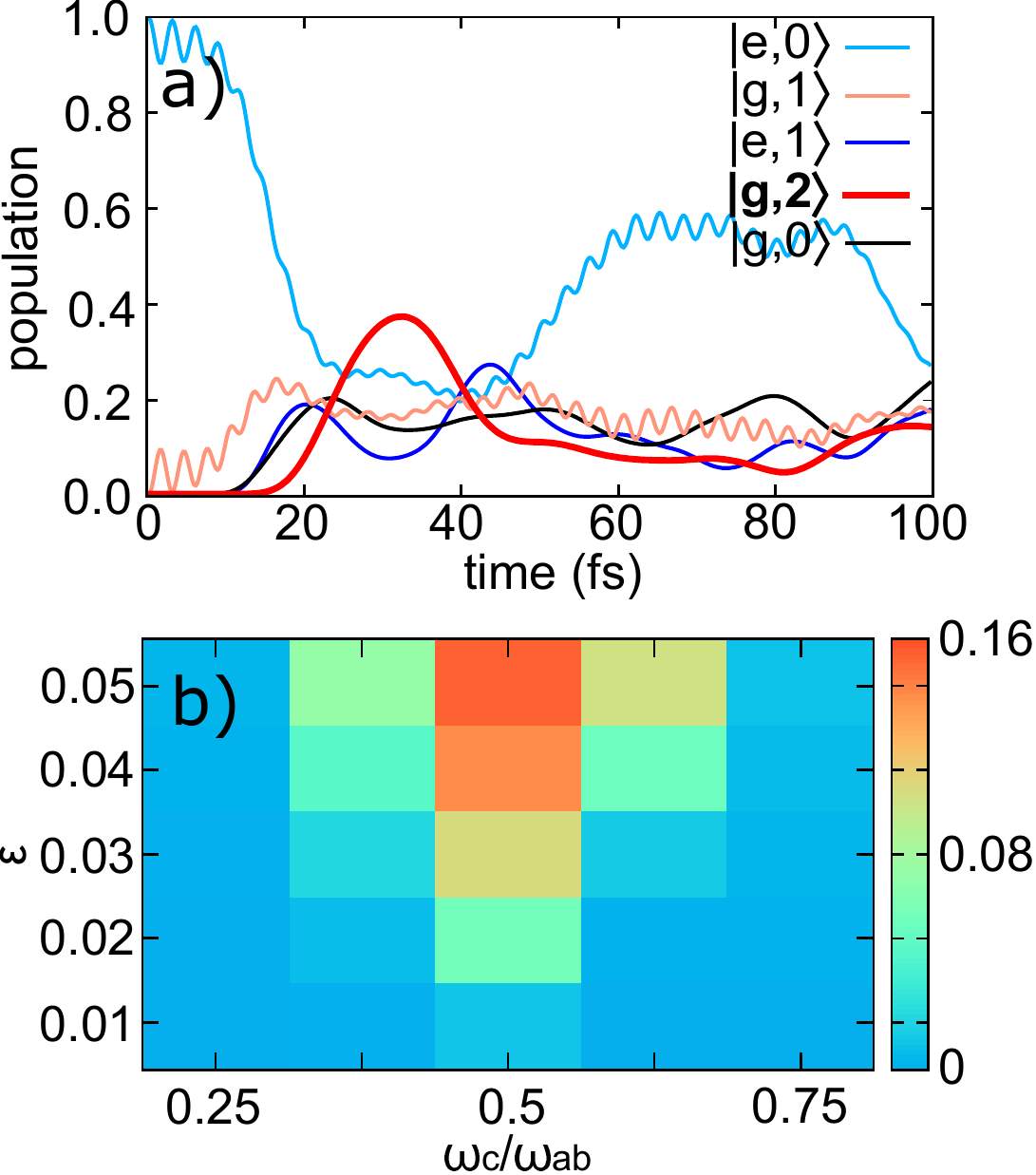}
\caption{a) Time-dependent adiabatic populations of photonic - electronic states for $\epsilon=0.04$ au ($g=0.21$ eV) and $\omega_{c}/\omega_{ab}=0.5$. b) Time average of two-photon population $\overline{P_{g,2}(t)}=\frac{1}{T}\int_{0}^{T}P_{g,2}(t)dt$ ($T=50$ fs) for different values of coupling strength $\epsilon$ and cavity frequency $\omega_{c}$.}
\label{pop}
\end{figure}
\end{center}

In Figure \ref{pop}a we present the adiabatic populations as a function of time. At short time, we can observe low amplitude and fast oscillations corresponding to off-resonance population transfer between $|e,0\rangle$ and $|g,1\rangle$ due to light-matter coupling. However, as the dynamics proceeds, those two states become resonant, and there is a fast decay of the initial state $|e,0\rangle$ first into $|g,1\rangle$, then into $|e,1\rangle$, and finally into $|g,0\rangle$ and $|g,2\rangle$ by the end of the isomerization. The population of $|g,2\rangle$ at 30 fs evidences the photon pair generation.
In Figure \ref{pop}b we see that if $\omega_{c}$ is too low or too large compared to $\omega_{ab}$, the state with two photons is not significantly populated. In the first case, the LIAC lies near the region in which the transition dipole moment is drastically reduced, suppressing light-matter coupling. In addition, the polaritonic PES near the LIAC is too steep, implying nuclear dynamics
that are too fast to be affected by the electronic-photonic coupling. In the second case, although light-matter coupling is not suppressed, the initial energy is not high enough to produce a nonadiabatic transition that would generate the second photon (see supplementary information Figure S3). Those inconveniences are overcome if $\omega_{c}$ is near half of $\omega_{ab}$ and the light-matter coupling is sufficiently strong so that a large population of the state $|g,2\rangle$ is produced. It should also be noticed that the cavity frequency does not have to be exactly half of the exciton frequency, as vibrations can account for the remaining energy to form the two photons. For a better understanding of the mechanism, we calculate the time dependent probability density for each adiabatic state:
\begin{equation}
\rho_{\kappa}(x,\phi,t)=\langle \Psi(t)| x,\phi,\kappa \rangle \langle x,\phi,\kappa | \Psi(t)\rangle,\ \kappa=g,e.
\end{equation}
\begin{center}
\begin{figure}[ht]
\includegraphics[scale=0.145]{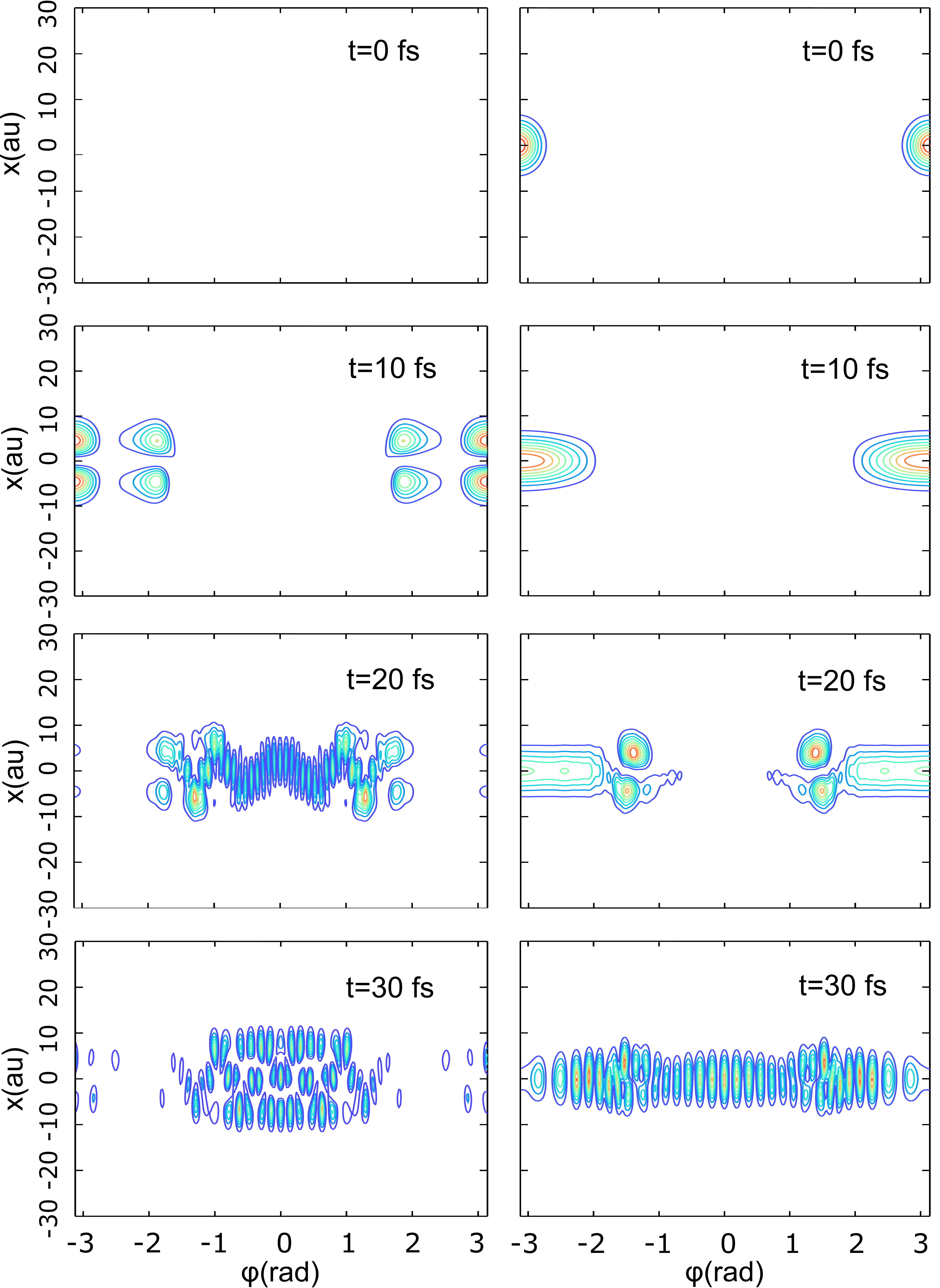}
\caption{Time dependent probability density along cavity ($x$) and nuclear ($\phi$) degrees of freedom for adiabatic electronic states $|g\rangle$ (left) and $|e\rangle$ (right). The largest number of photons in the wave function is given by the number of nodes along $x$ coordinate.}
\label{dynamics}
\end{figure}
\end{center}
Numerical simulations shown in Figure \ref{dynamics} support the mechanism proposed: at 0 fs the wavepacket corresponds to the Franck-Condon excitation of the {\it cis} molecular configuration with the cavity in the vacuum state $|e,0\rangle$. At 10 fs there is population of the state $|g,1\rangle$ due to strong light-matter coupling. Subsequently, population at the state $|e,1\rangle$ is created at 20 fs, indicating that nonadiabatic molecular dynamics has occurred. Finally, at $30$ fs we clearly notice formation of a wavepacket in the lower adiabatic state with two photons $|g,2\rangle$, in the {\it trans} configuration. Consistently, $30$ fs is the time when the population of the $|g,2\rangle$ state reaches its maximum value (see Figure \ref{pop}a). In a more comprehensive model, those populations would decay at long times due to cavity leakage (due to release of the down-converted photons) or vibrational relaxation (to vibrationally equilibrated populations that subsequently emit photons, as depicted in Figure \ref{polaritonPES}c). Even though we did not include those effects in our calculations, we do not expect these dissipative effects to affect the conclusions of this work as they occur within about 50 fs for organic microcavities at room temperature \cite{Raphael}, and the down-conversion process of interest happens before that. Other mechanisms that can be observed proceed as $|e,0\rangle \rightarrow |g,1\rangle \rightarrow |e,0\rangle$ (purely adiabatic dynamics and no photon down-conversion), $|e,0\rangle \rightarrow |g,1\rangle$ (adiabatic dynamics across one LIAC and no photon down-conversion), and $|e,0\rangle \rightarrow |g,0\rangle$ (purely nonadiabatic dynamics and no photon down-conversion).

\begin{center}
\begin{figure}
\includegraphics[scale=0.6]{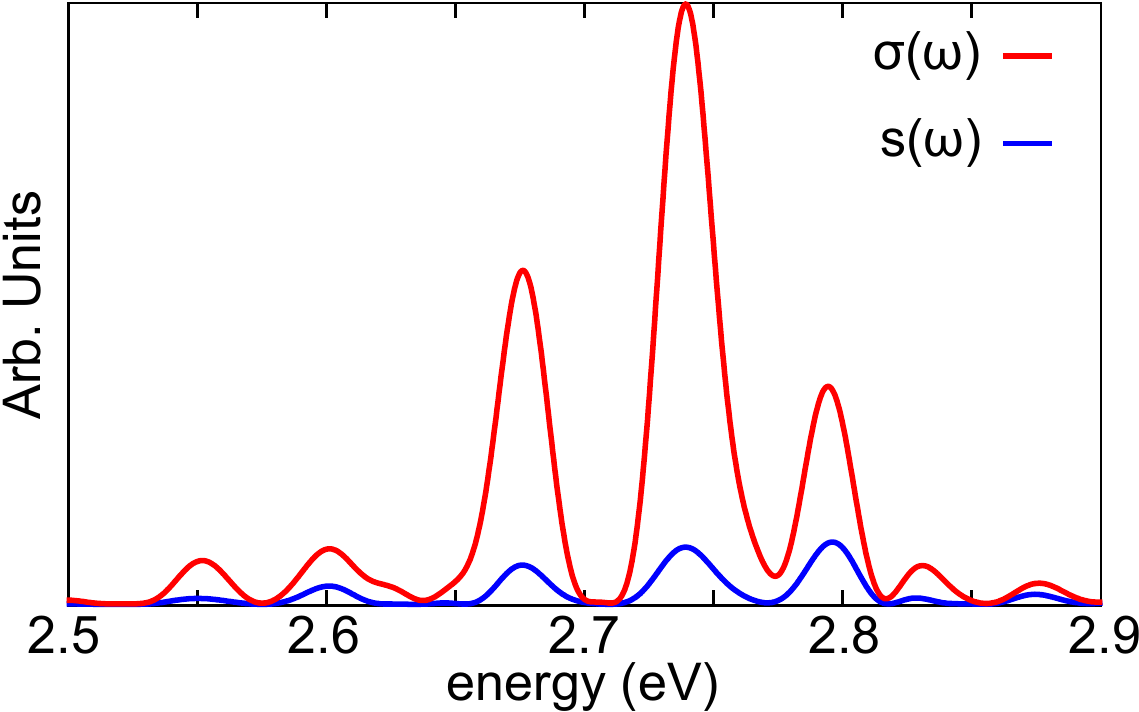}
\caption{Transition probabilities into polaritonic eigenstates due to direct optical excitation of the molecule via the dipole (red), and their two-photon character (blue).}
\label{spectrum}
\end{figure}
\end{center}

A complementary interpretation of the down-conversion mechanism can be provided from a time-independent perspective: for the molecule at the {\it cis} configuration, the upper polariton is mostly excitonic and accessible by means of a high-frequency photon. However, due to nonadiabatic couplings, the upper polariton is mixed with the lower polariton of the second excitation manifold at the {\it trans} configuration, which is accompanied by two photons. To further prove the two photon character of the eigenstates of $\hat{H}$ (Eq. 3) corresponding to the upper polariton PES, in Figure \ref{spectrum} we plot a variant of the absorption spectrum calculated as the spectral function of the initial wavepacket, and the spectral overlap of this wavepacket with the two-photon state $|2\rangle$. Spectral functions are computed as Fourier transforms of the following correlation functions (See supplementary information for further discussion of these calculations)\cite{Heller,Tannor}:
\begin{align}
&\sigma(\omega)=\int_{0}^{T}\int_{0}^{T} dt dt'\langle \Psi(t')| \Psi(t)\rangle e^{i(\omega+\omega_{0}) t}e^{-i(\omega+\omega_{0}) t'}\\
&s(\omega)=\int_{0}^{T}\int_{0}^{T}dt dt'\langle\Psi(t')|2\rangle\langle 2| \Psi(t)\rangle e^{i(\omega+\omega_{0}) t}e^{-i(\omega+\omega_{0}) t'},
\end{align} where $\omega_{0}$ is the energy of the ground state at the {\it cis} configuration and $T=100$ fs. The overlap of the two spectra evidences the two photon character of the polaritonic eigenstates in the absorption spectrum.

The presented photon down-conversion phenomenon constitutes a molecular analogue of the dynamical Casimir effect (DCE)\cite{Sanz}, where a cavity mode can be excited out of the vacuum by means of nonadiabatic variations of the mirrors in an optical microcavity \cite{Wilson,Souza}. Here, the kinetic energy is provided by the optical transition to the electronic excited state, forcing the nuclei to move out of the Frank-Condon region and undergo nonadiabatic molecular dynamics.

Experiments involving a single emitter strongly coupled to a confined electromagnetic field have been recently carried out by placing molecules on plasmonic nanocavities \cite{Chikkaraddy,Reithmaier,Ojambati,Santhosh,Park,Chikkaradd} and Fabry-Perot microcavities \cite{Sandoghdar}. However, one of the most common setups to achieve SC involves the use of a macroscopic amount of molecules. The electromagnetic field interacts with an ensemble of $N$ molecules to form 2 (upper and lower) polariton states, and $N-1$ mostly molecular (dark) states that mix weakly with light either due to disorder or vibrational motion \cite{Raphael,Feist2}. Under these conditions, collective light-matter coupling scales as $\sim\sqrt{N}\epsilon$, where $\epsilon$ is the individual light-matter coupling and $N$ is the number of particles in the microcavity.

To investigate the feasibility of achieving down-conversion using collective light-matter coupling, we performed calculations of two molecules embedded in an optical microcavity. Eq. 6 can be generalized to $N$ molecules \cite{Feist2,Huo}, assuming molecules interact identically with the cavity mode and have no direct electrostatic interaction between them. The Hamiltonian is a generalization of the Dicke Model \cite{Dicke} that includes the nuclear degrees of freedom:
\begin{equation}
\hat{H}= \sum_{i=1}^{N} \left(\hat{T}_{N,i} + \hat{H}_{el,i}\right) + \omega_{c}(\hat{a}^{\dagger}\hat{a}+1/2) +  \sum_{i=1}^{N} g(\phi_{i})(\hat{a}^{\dagger}+\hat{a})\hat{\sigma}_{x,i}.
\end{equation}
We study the two-photon generation at constant collective light-matter coupling for $N=1,2$ molecules (thus setting the individual light-matter coupling in each case at $\epsilon=0.04/\sqrt{N}$ au). We assume that only molecule 1 is initially excited, and calculate electronic state populations of each molecule at the {\it cis} ($|g_{C}\rangle$, $|e_{C}\rangle$) and {\it trans} ($|g_{T}\rangle$, $|e_{T}\rangle$) configurations using the projection operators 

\begin{equation}
\hat{P}_{\kappa,C}^{i}=\int d\phi_{i}|[\Theta(\phi_{i}-1.63)+\Theta(-\phi_{i}-1.63)]|\kappa_{i},\phi_{i}\rangle\langle \kappa_{i},\phi_{i}|
\end{equation} and
\begin{equation}
\hat{P}_{\kappa_{i},T}^{i}= 1 - \hat{P}_{\kappa_{i},C}^{i},
\end{equation} where $\hat{P}_{\kappa_{i},C}^{i}$ is the projector of the i-th molecule over the {\it cis} configuration and electronic state $\kappa_{i}$, and $\Theta(\phi)$ is the Heaviside step function. For instance, the probability of both molecules to be at the {\it cis} configuration in the ground state, i.e. state $|g_{C}g_{C}\rangle$ , is given by $\langle\hat{P}_{g,C}^{1}\hat{P}_{g,C}^{2} \rangle$. 

\begin{center}
\begin{figure}[ht]
\includegraphics[scale=0.55]{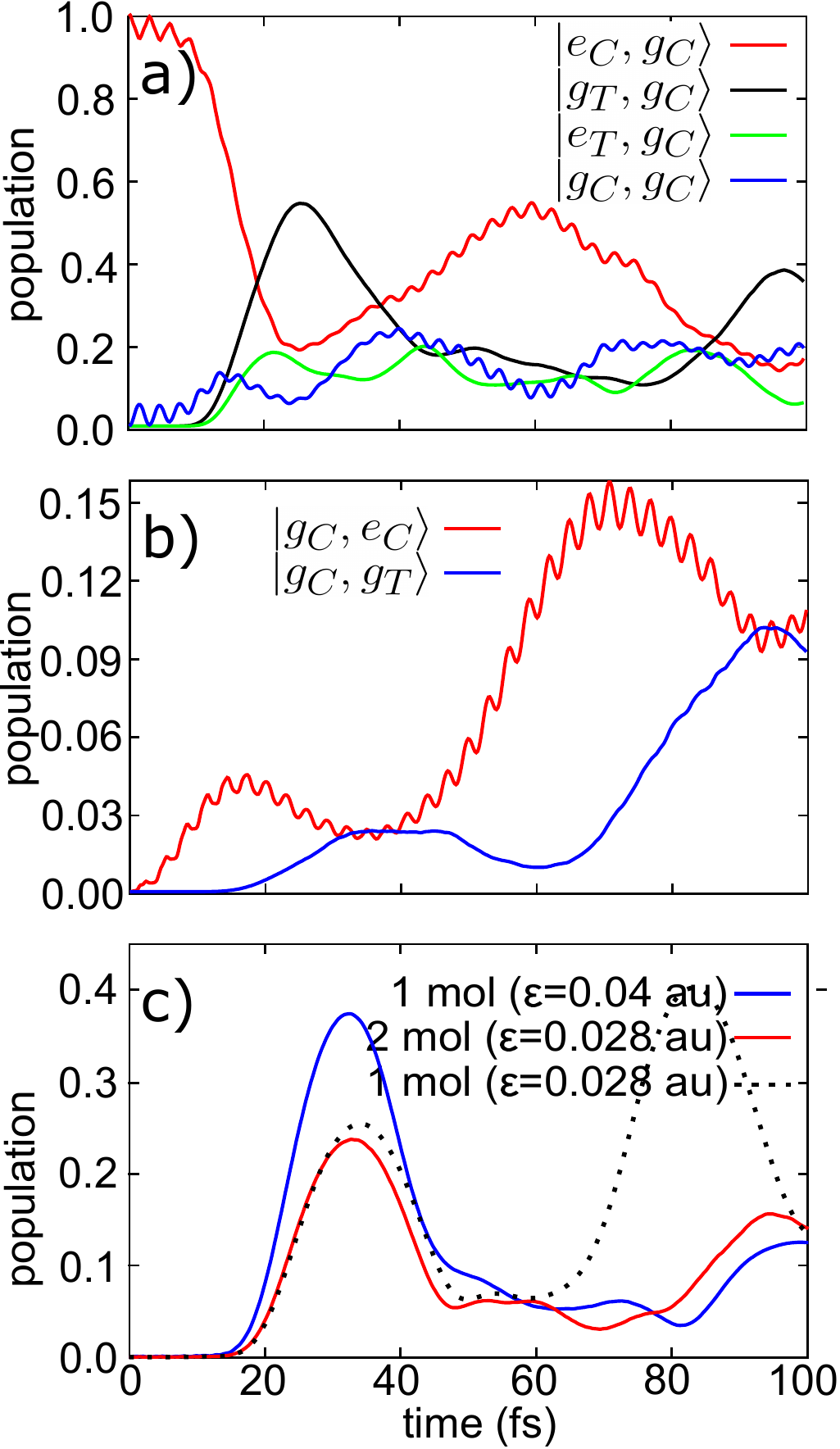}
\caption{Population dynamics for $\omega_{c}/\omega_{ab}=0.5$ and $\epsilon=0.03$ au, if only molecule 1 is initially excited. a) States with significant population during the first $100$ fs. b) Population of states in which molecule 2 undergoes isomerization reaction. Notice the difference in vertical scale. c) Comparison between populations of two-photon states for one and two molecules with the same {\it collective} light-matter coupling (blue {\it vs} red); and for two and one molecules with the same {\it individual} light-matter coupling (red \textit{vs} dashed black).}
\label{pops2mol}
\end{figure}
\end{center}

As can be observed in Figure \ref{pops2mol}a, most dynamics during the first $100$ fs involves the isomerization of the initially excited molecule, while the second molecule remains at the {\it cis} configuration in the ground state. This path resembles that of the single-molecule scenario, in which the excited molecule isomerizes in around $30$ fs, ending at the {\it trans} configuration in both the molecular ground and the excited state (see Figure \ref{dynamics}). After $40$ fs, the second molecule gets excited and also undergoes isomerization. However, this process seems to occur only if the first molecule is in its ground state at the {\it cis} configuration (Figure \ref{pops2mol}b). This can be understood by noticing that the molecules at the {\it cis} configuration are closer to resonance with the cavity mode. The individual coupling of each molecule with the cavity mode produces an effective coupling between them, causing the excitation of one molecule to be transferred to the other one (i.e $|e_{C}g_{C}\rangle \rightarrow |g_{C}e_{C}\rangle$), as have been reported in previous works \cite{Feist2,Huo}. Evidently this process is not very likely in our setup, and should not be observed in the limit where the single-molecule coupling is weak.

The mechanism involving two molecules can thus be summarized as follows:

\begin{enumerate}
\item Optical excitation of the first molecule ($|g_{C},g_{C}\rangle \rightarrow |e_{C},g_{C}\rangle$).
\item Isomerization of the first molecule. ($|e_{C},g_{C}\rangle \rightarrow |e_{T},g_{C}\rangle$ and $|e_{C},g_{C}\rangle \rightarrow |g_{T},g_{C}\rangle$). As in the single molecule scenario, this can produce zero, one or two photons.
\item Cavity-mediated excitation-energy transfer from the first to the second molecule at the {\it cis} configuration ($|e_{C}g_{C}\rangle \rightarrow |g_{C}e_{C}\rangle$).
\item Isomerization of the second molecule ($|g_{C},e_{C}\rangle \rightarrow |g_{C},e_{T}\rangle$ and $|g_{C},e_{C}\rangle \rightarrow |g_{C},g_{T}\rangle$), producing zero, one, or two photons.
\end{enumerate}

As one would expect based on the mechanism above, Figure \ref{pops2mol}c shows that having two molecules instead of one does not increase the likelihood of generation of two photons (see blue and red curves), so long as the {\it collective} light-matter coupling $\sqrt{N}\epsilon=0.04$ au remains the same. To reinforce the observation that the observed effect is essentially a single-molecule one, we notice that for fixed {\it individual} light-matter coupling $\epsilon=0.028$ au, the two-photon state populations for two and one molecules (see red and black dashed curves) is very similar at short times. The reason we do not observe significant collective effects is that the molecules at the initial configuration are not in resonance with the cavity, but become resonant as the isomerization proceeds. As  a consequence, initial excitation of one molecule can lead to efficient energy exchange with the cavity only after sufficient nuclear dynamics ensues, while the other molecule remains off resonant. This was observed to be true even if the initial excitation is shared by all molecules in superposition. In other words, the Rabi splittings for the relevant LIACs relevant for this process mainly depend on the single-molecule coupling even if many molecules are present. 

We believe that the fact that the cavity and the molecule are off-resonance at the initial configuration is not a vital characteristic of the down-conversion mechanism proposed here. If the molecules are resonant with the cavity at the Frank-Condon region, an excitation of the upper-polariton can still produce a molecular non-adiabatic transition that would generate a second excitation, which can afterwards become a second photon. The Rabi splitting in that case would be of a collective nature, and the two-photon generation would be possible if the isomerization is faster than the decay from the upper-polariton to the dark states. The non-adiabatic molecular dynamics associated to molecules which are resonant at the Frank-Condon configuration could not be observed for the molecule studied here, due to the particular characteristics of the PESs. Hence, the effectiveness of our down-conversion scheme in the collective regime is molecule-specific and will require additional investigations.

In summary, we have shown that strong light-matter coupling in conjunction with nonadiabatic molecular dynamics can lead to emerging nonlinear optical phenomena such as photon down-conversion. While much attention has been recently placed into the study novel chemical dynamics afforded by molecular polaritons, we wish to emphasize a complementary aspect of the problem that is equally rich and relevant: the use of molecular dynamics to generate new photonic phenomena. The elucidated effect operates at the single molecule SC regime, but we have provided plausible arguments that would allow to extend it to the collective regime, where many experiments are being currently performed.

J.B.P.S. and J.Y.Z. were funded with the NSF EAGER Award CHE 1836599.

\bibliography{ms}

\end{document}


\section{Computational Details}

For the numerical integration of the TDSE, we employ the MCTDH algorithm implemented in the Heidelberg package \citep{mctdh,mctdh2}. In the multi-set formulation, The wavefunction is expanded in a set of electronic states ($k=a,b$):

\begin{equation}
\langle x, \phi | \Psi(t)\rangle = \sum_{(k=a)}^{b}\psi^{(k)}(x,\phi,t) |k\rangle,
\end{equation} where the wavefunction $\psi^{(k)}(x,\phi,t)$ is represented as a linear combination of Hartree products, each one of them consisting of a product of so-called single-particle functions (SPFs), namely

\begin{equation}
\psi^{(k)}(x,\phi,t)=\sum^{n_{x}}_{j_{x}=1}\sum^{n_{\phi}}_{j_{\phi}=1}A^{(k)}_{j_{x}j_{\phi}}(t)\varphi^{(k)}_{j_{x}}(x,t)\times\varphi^{(k)}_{j_{\phi}}(\phi,t),
\end{equation} where $n_{x}$ and $n_{\phi}$ are the number of SPFs for the cavity and nuclear degrees of freedom respectively. For each SPF, in turn, a discrete variable representation (DVR) is used. 

For an accurate representation of the wavefunction, convergence of the dynamics with respect to the number of SPFs and number of DVR points for each degree of freedom must be ensured. We used $n_{x}=n_{\phi}=4$, and $199$ and $150$ grid points for the nuclear and photonic degrees of freedom respectively.

\begin{center}
\begin{figure}
\includegraphics[scale=0.8]{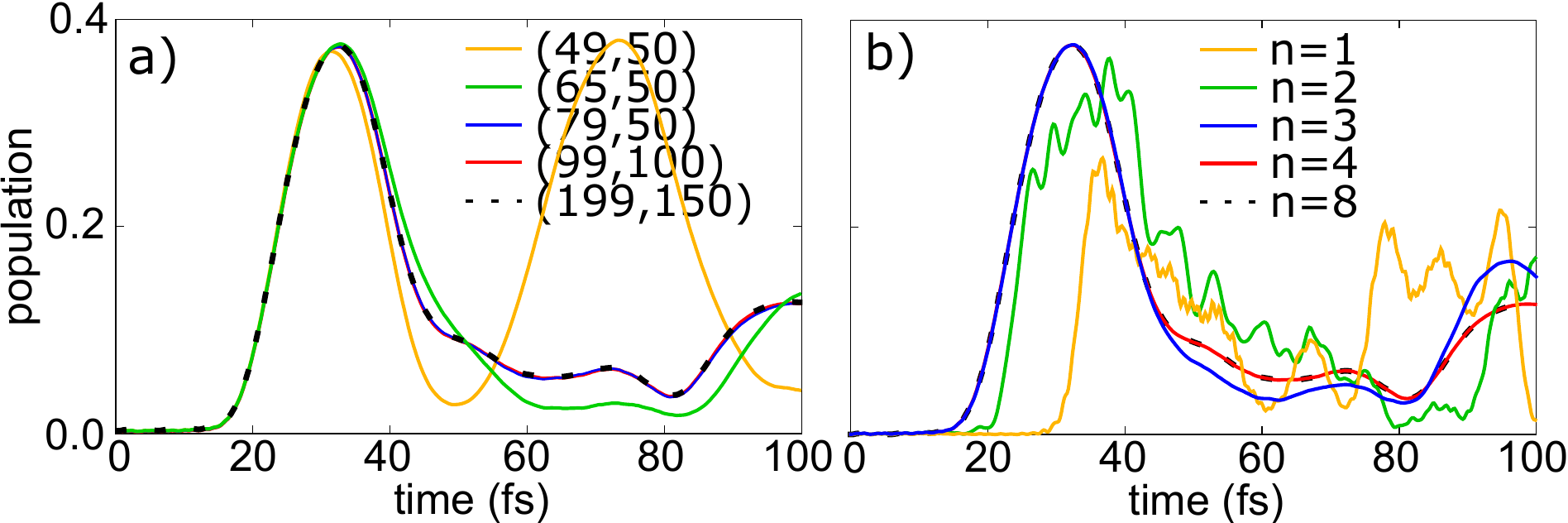}
\caption*{Figure S1: Convergence of simulations for 2-photon population for $\omega_{c}/\omega_{ab}=0.5$ and $\epsilon=0.04$. a) With respect to the number of grid points. b) With respect to the number of Single Particle Functions (SPFs).}
\end{figure}
\end{center}

\section{Dipole self-energy term}

In general, the dipole self-energy term $\frac{g(\phi)^{2}}{\omega_{c}}$ should be included in the Hamiltonian (Eq. 4) in order to ensure a bounded ground state of the polaritonic system \cite{Rubio3}. In Figure S2, we compare the results of the quantum dynamics simulations with and without such term in the Hamiltonian.

\begin{center}
\begin{figure} 
\includegraphics[scale=0.8]{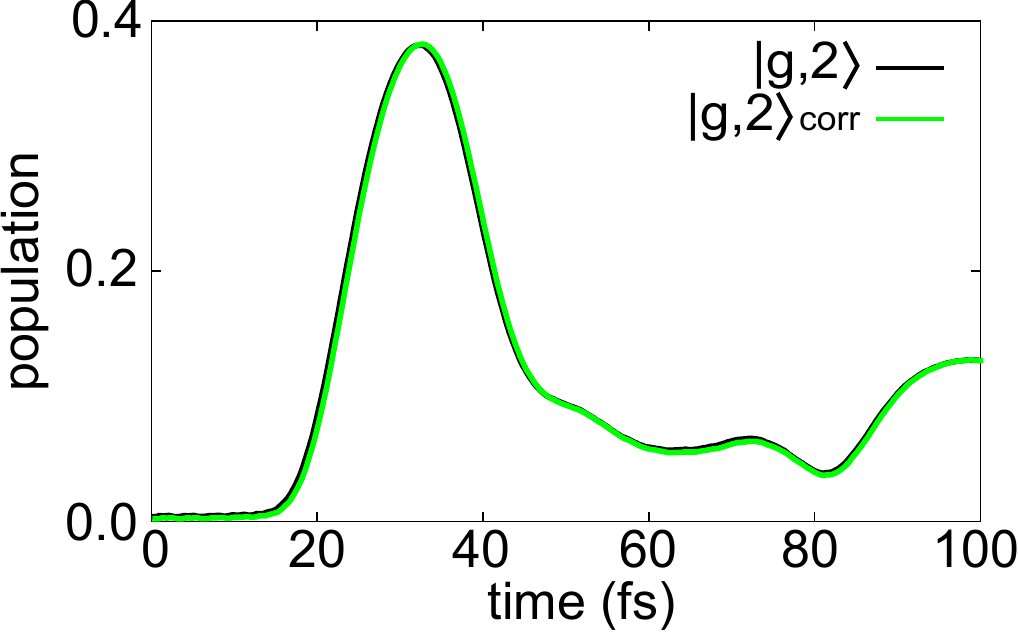}
\caption*{Figure S2: 2-photon population for $\omega_{c}/\omega_{ab}=0.5$ and $\epsilon=0.04$ with (green) and without (black) the correction dipole self-energy term.}
\end{figure}
\end{center}

It is evident that such term has almost no influence in the dynamics. This is because our model has no permanent dipole, light-matter coupling is not large enough, and the transition dipole moment is constant for most values of $\phi$ (see Figure 1). 

\section{Polaritonic energy surfaces for $\omega_{c}>\omega_{ab}/2$ and $\omega_{c}<\omega_{ab}/2$.}

The following figure shows the polaritonic PESs and the adiabatic populations for $\omega_{c}/\omega_{ab}\neq 0.5$. 

\begin{center}
\begin{figure}
\includegraphics[scale=0.7]{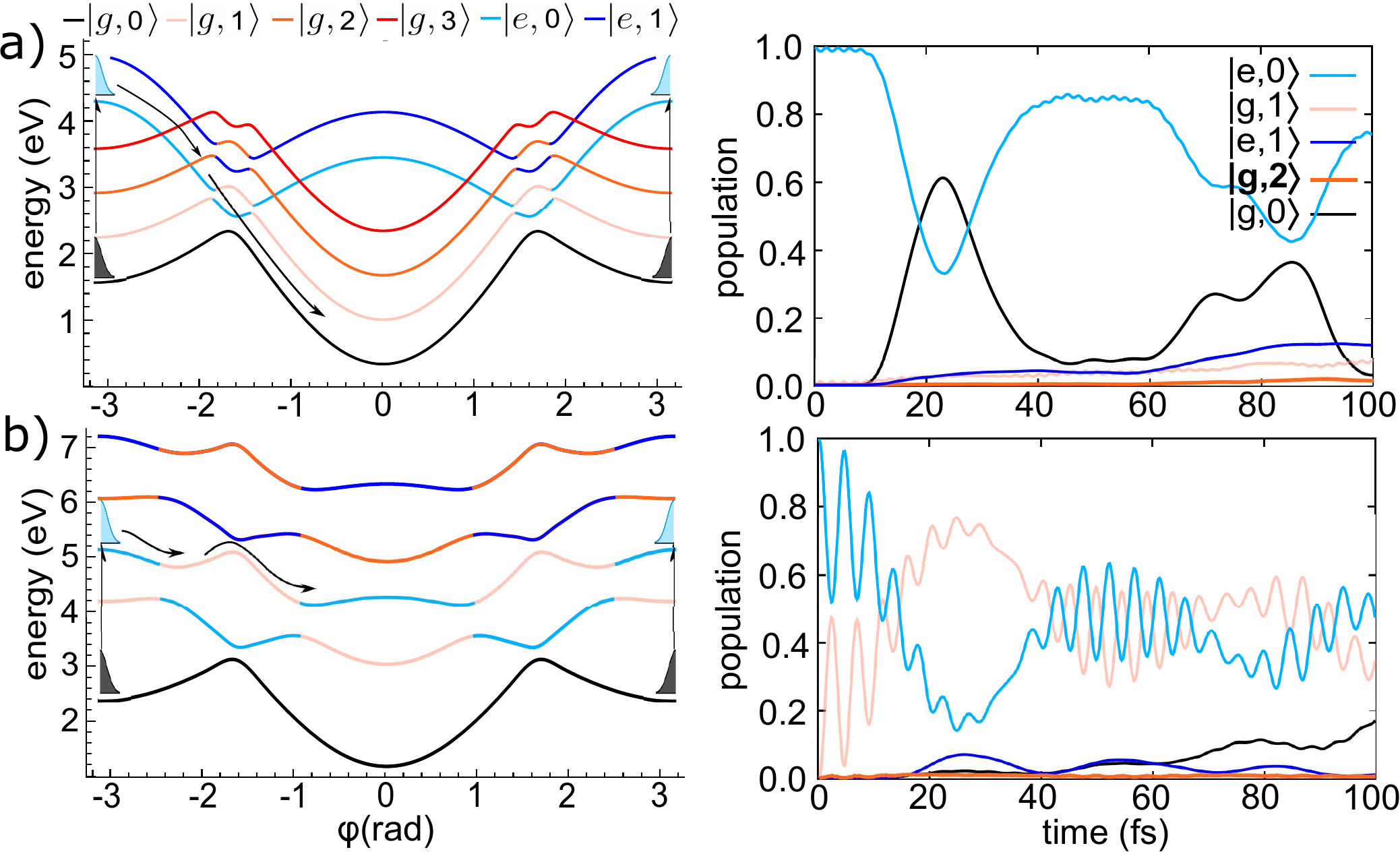}
\caption*{Figure S3: Polaritonic PES and population dynamics for numerical simulations at a) $\omega_{c}/\omega_{ab}=0.25$, $\epsilon=0.04$. b) $\omega_{c}/\omega_{ab}=0.75$, $\epsilon=0.04$. }
\end{figure}
\end{center}

For $\omega_{c}/\omega_{ab}<0.5$, the LIAC is reduced because it lies on the region in which transition dipole moment decreases. In addition, the polaritonic PES is steep, which favors a deleterious non-adiabatic transition through the LIAC. The featured dynamics proves this by showing no population of any photonic state. For $\omega_{c}/\omega_{ab}>0.5$, the LIAC is large enough to induce oscillation of excitonic and photonic state. However, the nuclear kinetic energy is not high enough to produce a molecular non-adiabatic transition.

\section{Spectral functions}

We are interested in the absorption spectrum of the polaritonic system via the molecular dipole moment $\hat{mu}$, but also in the two-photon character of the accessible polaritonic eigenstates of $\hat{H}$. The following procedure allows for such calculations.

The first-order perturbative state accessed by a weak laser field at frequency $\omega$ is proportional to \cite{Tannor}

\begin{equation}
    |\Psi^{(1)}(t,\omega)\rangle=\int_{0}^{t}dt' e^{-i\hat{H}(t-t')}\hat{\mu}e^{-i\omega t'}e^{-i\hat{H}t'}|\phi(0)\rangle,
\end{equation} where $|\phi(0)\rangle$ is the ground state of the system (molecule + cavity).

Given that $\hat{H}|\phi(0)\rangle = \omega_{0}|\phi(0)\rangle$, we have

\begin{equation}\label{eq:spectrum}
    |\Psi^{(1)}(t,\omega)\rangle=\int_{0}^{t}dt' e^{-i\hat{H}(t-t')}|\Psi(0)\rangle e^{-i(\omega+\omega_{0})t'},
\end{equation} where $|\Psi(0)\rangle=\hat{\mu}|\phi(0)\rangle$ is the initial state used in the computational simulations.

\subsection{Total absorption spectrum}

Using Eq. \ref{eq:spectrum}, the change of population due to the weak interaction with the EM field out of the cavity is given by 

\begin{equation} \label{spec}
   \sigma_{t}(\omega)\equiv\langle\Psi^{(1)}(t,\omega)|\Psi^{(1)}(t,\omega)\rangle = \int_{0}^{t}\int_{0}^{t}dt'dt'' \langle\Psi(t')|\Psi(t'')\rangle e^{i(\omega+\omega_{0})(t''-t')}.
  \end{equation}

Changing variables to $\tau=t''-t'$ and $\tau'=t''+t'$ we get

\begin{equation}
   \sigma_{t}(\omega) = t\int_{-t}^{t}d\tau \langle\Psi(0)|\Psi(\tau)\rangle e^{i(\omega+\omega_{0})\tau},
\end{equation} where  the integral in the right-hand side gives rise to the standard absorption spectrum:
\begin{equation}
    \int_{-t}^{t}d\tau \langle\Psi(0)|\Psi(\tau)\rangle e^{i(\omega+\omega_{0})\tau} = 2\sum_{n}|c_{n}|^{2}\frac{\sin{(\omega_{n}-(\omega+\omega_{0}))t}}{(\omega_{n}-(\omega+\omega_{0}))},
\end{equation} where $c_{n}=\langle n | \Psi(0)\rangle$ and $|n\rangle$ is and eigenstate of $\hat{H}$. In the Condon approximation, $c_{n}$ is the Frank-Condon factor.

Therefore, for long enough (yet finite) propagation times t, $\sigma_{t}(\omega)$ is proportional to the absorption spectrum

\begin{equation}
   \sigma_{t}(\omega) = 2t \sum_{n}|c_{n}|^{2}\frac{\sin{(\omega_{n}-(\omega+\omega_{0}))t}}{(\omega_{n}-(\omega+\omega_{0}))}.
\end{equation}

\subsection{Two photon character from absorption spectrum}

Note that Eq. \ref{spec} can also be written as 

\begin{equation}
     \sigma_{t}(\omega)=\sum_{i} \int_{0}^{t}\int_{0}^{t}dt'dt'' \langle\Psi(t')|i\rangle\langle i |\Psi(t'')\rangle e^{i(\omega+\omega_{0})(t''-t')}\equiv\sum_{i}s_{t}^{(i)}(\omega),
\end{equation} where $\{ |i\rangle\}$ is a complete basis set.

In this picture, the absorption spectrum $\sigma_{t}(\omega)$ is a sum of spectra $s_{t}^{(i)}(\omega)$. Based on this, for sufficiently large t values, a comparison  of both spectra $s_{t}^{(i)}(\omega)$ and $\sigma_{t}(\omega)$ gives us an estimation of the overlap of interest $\frac{s_{t}^{(i)}(\omega)}{\sigma_{t}(\omega)}=|\langle i | n \rangle|^{2}<1$ via a time-dependent approach. We use these ideas to calculate the two-photon character of eigenstates accessible by one-photon excitation, using $|i\rangle=|2\rangle$ (See Figure 5 in the manuscript). 

\bibliography{sup_info}